\documentclass[twocolumn]{openjournal}

\usepackage{lipsum}

\usepackage{xcolor}
\usepackage{textgreek}
\usepackage[utf8]{inputenc}
\usepackage[english]{babel}

\usepackage{hyperref}
\hypersetup{
    unicode, 
    colorlinks=true,
    linkcolor=linkcolor,
    citecolor=linkcolor,
    filecolor=linkcolor,
    urlcolor=linkcolor,
}
\usepackage{color,colortbl}
\definecolor{linkcolor}{rgb}{0.0,0.3,0.5}
\usepackage{tensind}
\tensordelimiter{?}
\DeclareGraphicsExtensions{.bmp,.png,.jpg,.pdf}
\usepackage{verbatim}
\usepackage[normalem]{ulem}
\usepackage{orcidlink}
\usepackage{soul}

\graphicspath{ {./figs/} }

\begin{document}
\title{On the maximum disk heating attributable to fuzzy dark matter}

\author{Chris Hamilton\orcidlink{0000-0002-5861-5687}}
\email{chamilton@ias.edu}
\affiliation{Institute for Advanced Study, Einstein Drive, Princeton, NJ 08540}

\begin{abstract}
Fuzzy dark matter (FDM) granulations would drive orbital transport of stars in galactic disks,
and in particular would produce roughly equal amounts of radial heating and radial migration.
However, observations suggest that heating has been much less efficient than migration in our Galaxy.
We argue that this decreases the amount of radial heating, $\mathcal{H}_\mathrm{FDM}$,  that 
can safely be attributed to FDM.
Consequently, lower bounds on the FDM particle mass $m$ derived through Galactic disk kinematics should be revised upwards; a rough estimate is $m \gtrsim 1.3\times 10^{-22} \mathrm{eV} \times
    [(\mathcal{H}_\mathrm{FDM}/\mathcal{H})/0.1]^{-1/2}$, where $\mathcal{H}$ is the total observed radial heating.
\end{abstract}


\maketitle

\section{Introduction}

Fuzzy dark matter (FDM) is a dark matter candidate that solves various small-scale problems often associated with $\Lambda$CDM cosmology \citep{hui2017ultralight}.  It proposes that the dark matter particle is of such low mass that its de Broglie wavelength is comparable to the 
size of galactic structures, typically $\sim 0.1-1$ kpc. 

The wave interference fringes or `granulations' that result from FDM would cause fluctuations in the gravitational potential of galaxies, driving orbital heating of the disk stars, i.e., increases in their radial actions $J_R$ and vertical actions $J_z$.
In particular, the smaller is the FDM particle mass $m$, the more vigorous is the orbital heating \citep{bar2019relaxation}. 
Recently \cite{chiang2023can,yang2024galactic} used this fact to place a lower bound on $m$. By assuming that all (vertical) heating of our Galaxy was caused by FDM, they concluded $m\gtrsim 0.4\times 10^{-22}$ eV.

However, little attention has been paid to the amount of radial \textit{migration} (change in guiding radius, or equivalently orbital angular momentum $J_\varphi$) that would naturally accompany this orbital heating. 
\cite{frankel2020keeping} used GAIA/APOGEE data to measure the radial migration $\mathcal{M} \equiv \mathrm{rms} \,\delta J_\varphi$ and radial heating $\mathcal{H} \equiv \mathrm{rms} \,\delta J_R$ that has occurred across our Galactic disk over the last $6$ Gyr.\footnote{Here, $\delta J_{\varphi, R}$ is the value of a given star's action $J_{\varphi, R}$ today minus the value it had $6$ Gyr ago. Then, rms $\delta J_{\varphi, R}$ is the root mean square of the $\delta J_{\varphi, R}$ values of all stars in the sample.}
They found that $\mathcal{H}/\mathcal{M} \approx 0.1$, i.e. migration is an order of magnitude more efficient than heating.
\cite{hamilton2024galactic} (hereafter HMT) claimed that such a small ratio could only be produced by a particular class of resonant spiral perturbations.
By contrast, FDM granulation-driven scattering is well-described as a series of uncorrelated, impulsive kicks \citep{bar2019relaxation,yang2024galactic,zupancic2024fuzzy}, and these would drive about as much heating as migration ($\mathcal{H}/\mathcal{M}\sim 1$), a ratio far in excess of that observed.

Here we use this discrepancy to show that the amount of allowed FDM-induced orbital heating in the Galaxy may be much lower than previously assumed, and that this could improve the (dynamical) lower bound on the FDM particle mass.  We present our calculation in \S\ref{sec:Calculation} and discuss the result in \S\ref{sec:Summary}.

\section{Calculation}
\label{sec:Calculation}

HMT argued that heating and migration due to many random, isotropic, impulsive kicks, of the sort provided by FDM, would be related by the following formula:
\begin{equation}
  \mathcal{H}_\mathrm{FDM}
   = \frac{f}{J_\varphi} \mathcal{M}_\mathrm{FDM}^2.
    \label{eqn:FDM_scaling}
\end{equation} Here $J_\varphi$ is the angular momentum of a circular orbit at the Galactocentric radius of interest, and $f$ is a pure number. 
Numerically, in a simple Solar-neighborhood-like model of a disk subjected to white noise Gaussian random fields we find
$f \approx 7 \pm 1$, though the result we will derive here is not too sensitive to $f$, as we will see.
The quadratic scaling
in (\ref{eqn:FDM_scaling}) arises because $J_R$ is quadratic in (radial) velocity while $J_\varphi$ is linear in (azimuthal) velocity \citep{Binney1988-zy}.

On the other hand, HMT also showed that resonant scattering by a particular class of transient spirals could drive heating that scaled roughly \textit{linearly} with migration:
\begin{equation}
    \mathcal{H}_\mathrm{res} =g\,  \mathcal{M}_\mathrm{res},
        \label{eqn:SS_scaling}
\end{equation}
where $g$ is another pure number. The linear scaling follows from the conservation of the Jacobi integral when a star interacts resonantly with a rigidly-rotating perturbation --- see \cite{Sellwood2002-lv}. 
HMT argued that the observed
$\mathcal{H}/\mathcal{M} \approx 0.1$ could only be explained if the main driver of orbital transport was resonant scattering following roughly the scaling (\ref{eqn:SS_scaling}), which obviously requires $g \lesssim 0.1$.
HMT found that a subset of reasonable spirals were indeed able to reproduce $g\approx 0.1$, although $g \ll 0.1$ was near-impossible achieve to except with very contrived spirals.

We now wish to know what constraints these scalings place on the FDM-driven heating.
In order to be as generous as possible to FDM, we will allow it to be one of only two mechanisms driving transport in the Galaxy, the other being resonant spiral scattering following (\ref{eqn:SS_scaling}) with a fixed `hotness' $g$ (so we are ignoring ISM clouds, infalling satellites, etc.) Thus we write 
\begin{eqnarray}
   && \mathcal{M} = 
    \mathcal{M}_\mathrm{FDM}
    +
    \mathcal{M}_\mathrm{res},
    \label{eqn:obs_migration_sum}
    \\
&&
    \mathcal{H} = 
    \mathcal{H}_\mathrm{FDM}
    +
    \mathcal{H}_\mathrm{res}.
        \label{eqn:obs_heating_sum}
\end{eqnarray}
Eliminating $\mathcal{M}_\mathrm{FDM}$, $\mathcal{M}_\mathrm{res}$ and $\mathcal{H}_\mathrm{res}$
from (\ref{eqn:FDM_scaling})-(\ref{eqn:obs_heating_sum}) 
gives a quadratic equation for $\mathcal{H}_\mathrm{FDM}$.
The relevant solution is
\begin{eqnarray}
  \mathcal{H}_\mathrm{FDM} \nonumber =&& \mathcal{H} - g \, \mathcal{M} + \frac{g^2 J_\varphi}{2f}
  \nonumber
  \\
  &&+\bigg[\frac{g^2 J_\varphi}{2f} \bigg( 
  \frac{g^2 J_\varphi}{2f} + 2[\mathcal{H} - g \, \mathcal{M}]
  \bigg)\bigg]^{1/2}.
\label{eqn:rmsdJRFDM}
\end{eqnarray}
(The other solution is nearly always unphysical because it corresponds to negative $\mathcal{M}_\mathrm{FDM}$).

\begin{figure}
    \centering
    \includegraphics[width=0.45\textwidth]{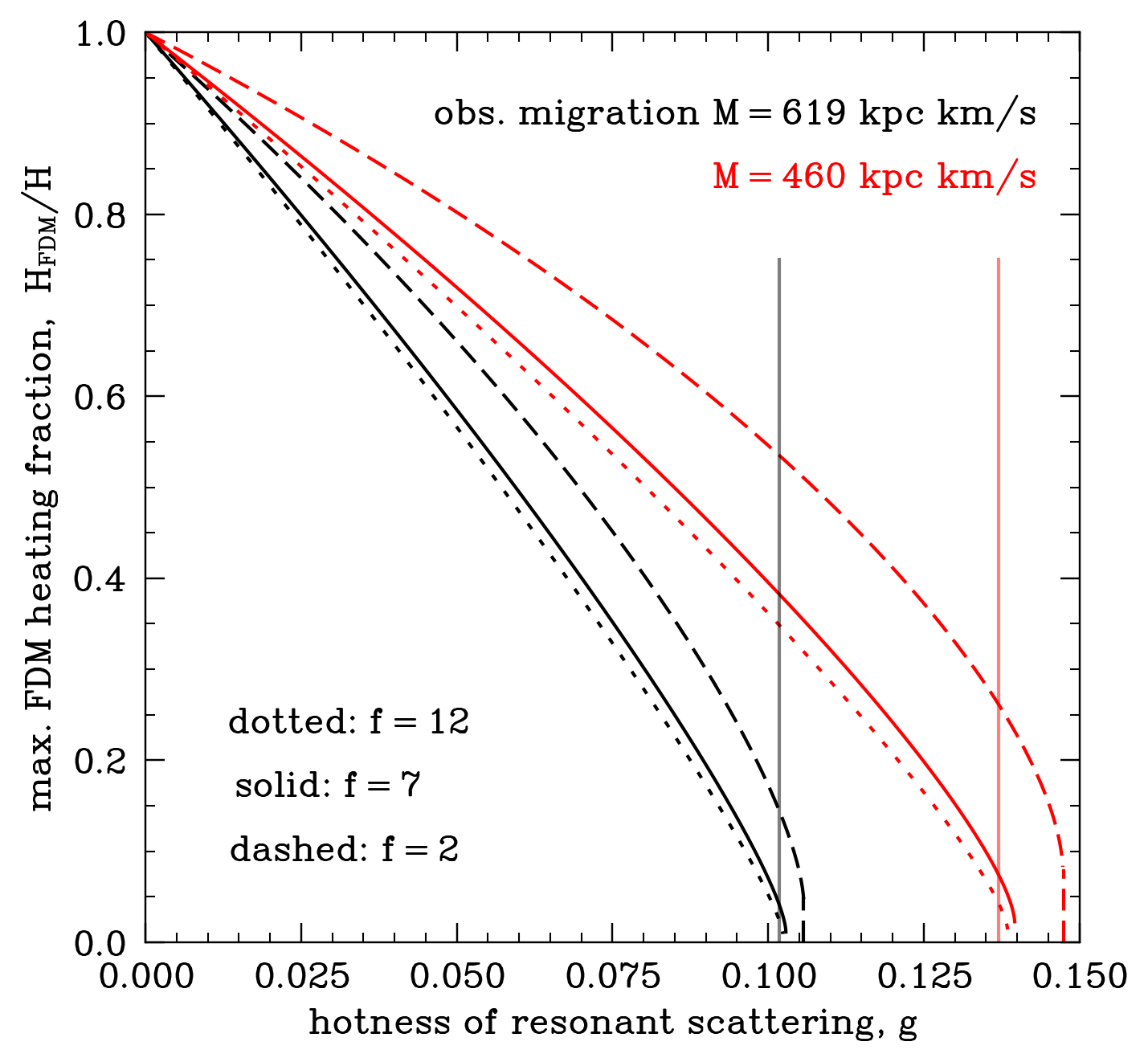}
    \caption{Plot of the maximum fraction of radial heating attributable to FDM, $\mathcal{H}_\mathrm{FDM}/\mathcal{H}$ (see equation (\ref{eqn:rmsdJRFDM})), as a function of the `hotness' of resonant spiral transport $g$ (equation (\ref{eqn:SS_scaling})).
    The black solid curve is for the fiducial values $J_\varphi = 1760$ kpc km s$^{-1}$, $\mathcal{M} = 619$ kpc km s$^{-1}$,
    $\mathcal{H} = 63$ kpc km s$^{-1}$ and 
$f=7$.  The red curves assume a reduced observed radial migration,
$\mathcal{M} = 460$ kpc km s$^{-1}$, and the dashed (dotted) curves assume $f = 2$ ($12$). Vertical lines correspond to $g = \mathcal{H}/\mathcal{M}$.}
    \label{fig:illustration1}
\end{figure}

Equation (\ref{eqn:rmsdJRFDM}) tells us the maximum amount of heating that we may attribute to FDM before falling foul of the observations.
In Figure \ref{fig:illustration1}, with a black solid line we plot the ratio $\mathcal{H}_\mathrm{FDM}/\mathcal{H}$ from equation (\ref{eqn:rmsdJRFDM}) as a function of $g$,
for fiducial Solar neighborhood values $J_\varphi = 1760$ kpc km s$^{-1}$, $\mathcal{M} = 619$ kpc km s$^{-1}$,
    $\mathcal{H} = 63$ kpc km s$^{-1}$ \citep{frankel2020keeping} and $f=7$.
    The vertical black line shows $g=\mathcal{H}/\mathcal{M} = 0.102$.
    We see that in the limit $g\to 0$, which corresponds to 
    resonant transport that is completely cold,
the ratio 
$\mathcal{H}_\mathrm{FDM}/\mathcal{H} \to 1$, i.e. 
FDM can be responsible for all the observed heating.
However, in the much more realistic case that $g$ approaches $\approx 0.1$,
the fraction of heating attributable to FDM decreases significantly.
For instance, $g=0.095$ corresponds to a maximum FDM heating fraction of $\mathcal{H}_\mathrm{FDM}/\mathcal{H} \approx 0.1$.


    The most  uncertain parameters we used in this calculation are the 
 observed radial migration $\mathcal{M}$
 (which is difficult to measure accurately)
 and the dimensionless number $f$ (which is a theoretical fudge factor fit to simplified simulations).
    To test the dependence of our results on these parameters, with the red curves in Figure \ref{fig:illustration1} we show the results of reducing $\mathcal{M}$ by $25\%$,
and with the dashed (dotted) curves we perform the same calculations assuming $f = 2$ ($f=12$). 
The reduction in $\mathcal{M}$ in particular has a significant impact on the result: 
with $f=7$ (red solid line), 
a $g$ value of $0.095$ now allows for a FDM heating fraction as large as $\mathcal{H}_\mathrm{FDM}/\mathcal{H} \approx 0.45$. 

    Interestingly, in all cases there are valid solutions 
    $\mathcal{H}_\mathrm{FDM}/\mathcal{H}$ for values of $g$ slightly \textit{larger} than $\mathcal{H}/\mathcal{M}$ (see the vertical lines in Figure \ref{fig:illustration1}). This would not be possible if $100\%$ of transport was resonant scattering following equation (\ref{eqn:SS_scaling}). It is possible here because the quadratic scaling in equation (\ref{eqn:FDM_scaling}) allows for small amounts of FDM-driven migration without a significant price being paid in terms of heating. 

\section{Discussion}
\label{sec:Summary}

We can convert our result on the maximum FDM heating fraction
into a rough lower bound on the FDM particle mass $m$ as follows.
First, \cite{chiang2023can,yang2024galactic} found that FDM granulations heat the Galactic disk \textit{vertically} at a rate proportional to 
$m^{-\alpha}$; simple theory predicts $\alpha = 3$, but simulations show something closer to $\alpha=2$,
and we will adopt the latter here.
Next, by assuming that $100\%$ of observed vertical heating was due to FDM, these authors
concluded $m\gtrsim 0.4\times 10^{-22} \,\mathrm{eV}$.
Now, in the approximation of isotropic impulsive scattering, radial and vertical heating rates
should be proportional to each other;
and since the
ratio of radial to vertical velocity dispersions in the Galaxy is close to that expected from isotropic impulsive scattering \citep{lacey1985massive,mackereth2019dynamical,ludlow2021spurious}, we can assume that the results of \cite{chiang2023can,yang2024galactic} correspond also to a \textit{radial} heating fraction due to FDM of $100\%$.
With these assumptions we can write $\mathcal{H}_\mathrm{FDM}/\mathcal{H} = [m/ 0.4\times 10^{-22} \,\mathrm{eV}]^{-2}$,
which can be inverted to give
\begin{equation}
    m \gtrsim 1.3\times 10^{-22} \mathrm{eV} \times
    \left(\frac{\mathcal{H}_\mathrm{FDM}/\mathcal{H}}{0.1}\right)^{-1/2}.
    \label{eqn:massbound}
\end{equation}
From Figure \ref{fig:illustration1} we know that the ratio $\mathcal{H}_\mathrm{FDM}/\mathcal{H}$
depends most strongly on the parameter $g$,
which sets the hotness of resonant spiral transport.
While we do not know $g$ a priori, 
the simulations of HMT show that only very contrived spirals are capable of producing $g \ll 0.1$. The closer $g$ is to $0.1$, the less heating is attributable to FDM, and the stronger is the lower bound on $m$. For example, with fiducial parameters (solid black line in Figure \ref{fig:illustration1}), $g\approx 0.095$ would give rise to $m \gtrsim 1.3\times 10^{-22}$ eV (equation (\ref{eqn:massbound})),
a factor $\sim 3$ higher than the bound from \cite{chiang2023can,yang2024galactic}.





We emphasize that stronger bounds on $m$ have been derived from, e.g.,
observations of the Lyman-alpha forest \citep{rogers2021strong} and ultra-faint dwarf galaxies \citep{dalal2022excluding}. 
However, all such bounds are uncertain and necessarily model-dependent, and our
Galactic disk heating and migration study provides an independent bound. 
Also, our arguments have been rather conservative, and including any additional physical ingredients in our model (such as gas clouds) would tend to limit the FDM parameter space even further.
We expect that our bound could be significantly improved by extending the analysis (both theory and data) to the vertical direction, by performing age- and radius-dependent calculations \citep{lian2022quantifying}, and/or by measuring migration and heating in other disk galaxies \citep{magrini2016metallicity}.
The arguments promoted here can also be used to guide careful numerical simulations.


\section*{Acknowledgments}

I thank J.-B. Fouvry, G. Tomaselli and M. Zaldarriaga for helpful comments on the manuscript.
C.H. is supported by the John N. Bahcall Fellowship Fund and the Sivian Fund at the Institute for Advanced Study.

\bibliographystyle{apsrev4-1}

\bibliography{oja_template}

\begin{thebibliography}{16}%
\makeatletter
\providecommand \@ifxundefined [1]{%
 \@ifx{#1\undefined}
}%
\providecommand \@ifnum [1]{%
 \ifnum #1\expandafter \@firstoftwo
 \else \expandafter \@secondoftwo
 \fi
}%
\providecommand \@ifx [1]{%
 \ifx #1\expandafter \@firstoftwo
 \else \expandafter \@secondoftwo
 \fi
}%
\providecommand \natexlab [1]{#1}%
\providecommand \enquote  [1]{``#1''}%
\providecommand \bibnamefont  [1]{#1}%
\providecommand \bibfnamefont [1]{#1}%
\providecommand \citenamefont [1]{#1}%
\providecommand \href@noop [0]{\@secondoftwo}%
\providecommand \href [0]{\begingroup \@sanitize@url \@href}%
\providecommand \@href[1]{\@@startlink{#1}\@@href}%
\providecommand \@@href[1]{\endgroup#1\@@endlink}%
\providecommand \@sanitize@url [0]{\catcode `\\12\catcode `\$12\catcode `\&12\catcode `\#12\catcode `\^12\catcode `\_12\catcode `\%12\relax}%
\providecommand \@@startlink[1]{}%
\providecommand \@@endlink[0]{}%
\providecommand \url  [0]{\begingroup\@sanitize@url \@url }%
\providecommand \@url [1]{\endgroup\@href {#1}{\urlprefix }}%
\providecommand \urlprefix  [0]{URL }%
\providecommand \Eprint [0]{\href }%
\providecommand \doibase [0]{http://dx.doi.org/}%
\providecommand \selectlanguage [0]{\@gobble}%
\providecommand \bibinfo  [0]{\@secondoftwo}%
\providecommand \bibfield  [0]{\@secondoftwo}%
\providecommand \translation [1]{[#1]}%
\providecommand \BibitemOpen [0]{}%
\providecommand \bibitemStop [0]{}%
\providecommand \bibitemNoStop [0]{.\EOS\space}%
\providecommand \EOS [0]{\spacefactor3000\relax}%
\providecommand \BibitemShut  [1]{\csname bibitem#1\endcsname}%
\let\auto@bib@innerbib\@empty
\bibitem [{\citenamefont {Hui}\ \emph {et~al.}(2017)\citenamefont {Hui}, \citenamefont {Ostriker}, \citenamefont {Tremaine},\ and\ \citenamefont {Witten}}]{hui2017ultralight}%
  \BibitemOpen
  \bibfield  {author} {\bibinfo {author} {\bibfnamefont {L.}~\bibnamefont {Hui}}, \bibinfo {author} {\bibfnamefont {J.~P.}\ \bibnamefont {Ostriker}}, \bibinfo {author} {\bibfnamefont {S.}~\bibnamefont {Tremaine}}, \ and\ \bibinfo {author} {\bibfnamefont {E.}~\bibnamefont {Witten}},\ }\href@noop {} {\bibfield  {journal} {\bibinfo  {journal} {Physical Review D}\ }\textbf {\bibinfo {volume} {95}},\ \bibinfo {pages} {043541} (\bibinfo {year} {2017})}\BibitemShut {NoStop}%
\bibitem [{\citenamefont {Bar-Or}\ \emph {et~al.}(2019)\citenamefont {Bar-Or}, \citenamefont {Fouvry},\ and\ \citenamefont {Tremaine}}]{bar2019relaxation}%
  \BibitemOpen
  \bibfield  {author} {\bibinfo {author} {\bibfnamefont {B.}~\bibnamefont {Bar-Or}}, \bibinfo {author} {\bibfnamefont {J.-B.}\ \bibnamefont {Fouvry}}, \ and\ \bibinfo {author} {\bibfnamefont {S.}~\bibnamefont {Tremaine}},\ }\href@noop {} {\bibfield  {journal} {\bibinfo  {journal} {The Astrophysical Journal}\ }\textbf {\bibinfo {volume} {871}},\ \bibinfo {pages} {28} (\bibinfo {year} {2019})}\BibitemShut {NoStop}%
\bibitem [{\citenamefont {Chiang}\ \emph {et~al.}(2023)\citenamefont {Chiang}, \citenamefont {Ostriker},\ and\ \citenamefont {Schive}}]{chiang2023can}%
  \BibitemOpen
  \bibfield  {author} {\bibinfo {author} {\bibfnamefont {B.~T.}\ \bibnamefont {Chiang}}, \bibinfo {author} {\bibfnamefont {J.~P.}\ \bibnamefont {Ostriker}}, \ and\ \bibinfo {author} {\bibfnamefont {H.-Y.}\ \bibnamefont {Schive}},\ }\href@noop {} {\bibfield  {journal} {\bibinfo  {journal} {Monthly Notices of the Royal Astronomical Society}\ }\textbf {\bibinfo {volume} {518}},\ \bibinfo {pages} {4045} (\bibinfo {year} {2023})}\BibitemShut {NoStop}%
\bibitem [{\citenamefont {Yang}\ \emph {et~al.}(2024)\citenamefont {Yang}, \citenamefont {Chiang}, \citenamefont {Su}, \citenamefont {Schive}, \citenamefont {Chiueh},\ and\ \citenamefont {Ostriker}}]{yang2024galactic}%
  \BibitemOpen
  \bibfield  {author} {\bibinfo {author} {\bibfnamefont {H.-Y.}\ \bibnamefont {Yang}}, \bibinfo {author} {\bibfnamefont {B.~T.}\ \bibnamefont {Chiang}}, \bibinfo {author} {\bibfnamefont {G.-M.}\ \bibnamefont {Su}}, \bibinfo {author} {\bibfnamefont {H.-Y.}\ \bibnamefont {Schive}}, \bibinfo {author} {\bibfnamefont {T.}~\bibnamefont {Chiueh}}, \ and\ \bibinfo {author} {\bibfnamefont {J.~P.}\ \bibnamefont {Ostriker}},\ }\href@noop {} {\bibfield  {journal} {\bibinfo  {journal} {Monthly Notices of the Royal Astronomical Society}\ }\textbf {\bibinfo {volume} {530}},\ \bibinfo {pages} {129} (\bibinfo {year} {2024})}\BibitemShut {NoStop}%
\bibitem [{\citenamefont {Frankel}\ \emph {et~al.}(2020)\citenamefont {Frankel}, \citenamefont {Sanders}, \citenamefont {Ting},\ and\ \citenamefont {Rix}}]{frankel2020keeping}%
  \BibitemOpen
  \bibfield  {author} {\bibinfo {author} {\bibfnamefont {N.}~\bibnamefont {Frankel}}, \bibinfo {author} {\bibfnamefont {J.}~\bibnamefont {Sanders}}, \bibinfo {author} {\bibfnamefont {Y.-S.}\ \bibnamefont {Ting}}, \ and\ \bibinfo {author} {\bibfnamefont {H.-W.}\ \bibnamefont {Rix}},\ }\href@noop {} {\bibfield  {journal} {\bibinfo  {journal} {The Astrophysical Journal}\ }\textbf {\bibinfo {volume} {896}},\ \bibinfo {pages} {15} (\bibinfo {year} {2020})}\BibitemShut {NoStop}%
\bibitem [{\citenamefont {Hamilton}\ \emph {et~al.}(2024)\citenamefont {Hamilton}, \citenamefont {Modak},\ and\ \citenamefont {Tremaine}}]{hamilton2024galactic}%
  \BibitemOpen
  \bibfield  {author} {\bibinfo {author} {\bibfnamefont {C.}~\bibnamefont {Hamilton}}, \bibinfo {author} {\bibfnamefont {S.}~\bibnamefont {Modak}}, \ and\ \bibinfo {author} {\bibfnamefont {S.}~\bibnamefont {Tremaine}},\ }\href@noop {} {\bibfield  {journal} {\bibinfo  {journal} {arXiv preprint arXiv:2411.08944}\ } (\bibinfo {year} {2024})}\BibitemShut {NoStop}%
\bibitem [{\citenamefont {Zupancic}\ and\ \citenamefont {Widrow}(2024)}]{zupancic2024fuzzy}%
  \BibitemOpen
  \bibfield  {author} {\bibinfo {author} {\bibfnamefont {B.}~\bibnamefont {Zupancic}}\ and\ \bibinfo {author} {\bibfnamefont {L.~M.}\ \bibnamefont {Widrow}},\ }\href@noop {} {\bibfield  {journal} {\bibinfo  {journal} {Monthly Notices of the Royal Astronomical Society}\ }\textbf {\bibinfo {volume} {527}},\ \bibinfo {pages} {6189} (\bibinfo {year} {2024})}\BibitemShut {NoStop}%
\bibitem [{\citenamefont {Binney}\ and\ \citenamefont {Lacey}(1988)}]{Binney1988-zy}%
  \BibitemOpen
  \bibfield  {author} {\bibinfo {author} {\bibfnamefont {J.}~\bibnamefont {Binney}}\ and\ \bibinfo {author} {\bibfnamefont {C.}~\bibnamefont {Lacey}},\ }\href@noop {} {\bibfield  {journal} {\bibinfo  {journal} {\mnras}\ }\textbf {\bibinfo {volume} {230}},\ \bibinfo {pages} {597} (\bibinfo {year} {1988})}\BibitemShut {NoStop}%
\bibitem [{\citenamefont {Sellwood}\ and\ \citenamefont {Binney}(2002)}]{Sellwood2002-lv}%
  \BibitemOpen
  \bibfield  {author} {\bibinfo {author} {\bibfnamefont {J.~A.}\ \bibnamefont {Sellwood}}\ and\ \bibinfo {author} {\bibfnamefont {J.~J.}\ \bibnamefont {Binney}},\ }\href@noop {} {\bibfield  {journal} {\bibinfo  {journal} {\mnras}\ }\textbf {\bibinfo {volume} {336}},\ \bibinfo {pages} {785} (\bibinfo {year} {2002})}\BibitemShut {NoStop}%
\bibitem [{\citenamefont {Lacey}\ and\ \citenamefont {Ostriker}(1985)}]{lacey1985massive}%
  \BibitemOpen
  \bibfield  {author} {\bibinfo {author} {\bibfnamefont {C.~G.}\ \bibnamefont {Lacey}}\ and\ \bibinfo {author} {\bibfnamefont {J.~P.}\ \bibnamefont {Ostriker}},\ }\href@noop {} {\bibfield  {journal} {\bibinfo  {journal} {Astrophysical Journal, Part 1 (ISSN 0004-637X), vol. 299, Dec. 15, 1985, p. 633-652.}\ }\textbf {\bibinfo {volume} {299}},\ \bibinfo {pages} {633} (\bibinfo {year} {1985})}\BibitemShut {NoStop}%
\bibitem [{\citenamefont {Mackereth}\ \emph {et~al.}(2019)\citenamefont {Mackereth}, \citenamefont {Bovy}, \citenamefont {Leung}, \citenamefont {Schiavon}, \citenamefont {Trick}, \citenamefont {Chaplin}, \citenamefont {Cunha}, \citenamefont {Feuillet}, \citenamefont {Majewski}, \citenamefont {Martig} \emph {et~al.}}]{mackereth2019dynamical}%
  \BibitemOpen
  \bibfield  {author} {\bibinfo {author} {\bibfnamefont {J.~T.}\ \bibnamefont {Mackereth}}, \bibinfo {author} {\bibfnamefont {J.}~\bibnamefont {Bovy}}, \bibinfo {author} {\bibfnamefont {H.~W.}\ \bibnamefont {Leung}}, \bibinfo {author} {\bibfnamefont {R.~P.}\ \bibnamefont {Schiavon}}, \bibinfo {author} {\bibfnamefont {W.~H.}\ \bibnamefont {Trick}}, \bibinfo {author} {\bibfnamefont {W.~J.}\ \bibnamefont {Chaplin}}, \bibinfo {author} {\bibfnamefont {K.}~\bibnamefont {Cunha}}, \bibinfo {author} {\bibfnamefont {D.~K.}\ \bibnamefont {Feuillet}}, \bibinfo {author} {\bibfnamefont {S.~R.}\ \bibnamefont {Majewski}}, \bibinfo {author} {\bibfnamefont {M.}~\bibnamefont {Martig}},  \emph {et~al.},\ }\href@noop {} {\bibfield  {journal} {\bibinfo  {journal} {Monthly Notices of the Royal Astronomical Society}\ }\textbf {\bibinfo {volume} {489}},\ \bibinfo {pages} {176} (\bibinfo {year} {2019})}\BibitemShut {NoStop}%
\bibitem [{\citenamefont {Ludlow}\ \emph {et~al.}(2021)\citenamefont {Ludlow}, \citenamefont {Fall}, \citenamefont {Schaye},\ and\ \citenamefont {Obreschkow}}]{ludlow2021spurious}%
  \BibitemOpen
  \bibfield  {author} {\bibinfo {author} {\bibfnamefont {A.~D.}\ \bibnamefont {Ludlow}}, \bibinfo {author} {\bibfnamefont {S.~M.}\ \bibnamefont {Fall}}, \bibinfo {author} {\bibfnamefont {J.}~\bibnamefont {Schaye}}, \ and\ \bibinfo {author} {\bibfnamefont {D.}~\bibnamefont {Obreschkow}},\ }\href@noop {} {\bibfield  {journal} {\bibinfo  {journal} {Monthly Notices of the Royal Astronomical Society}\ }\textbf {\bibinfo {volume} {508}},\ \bibinfo {pages} {5114} (\bibinfo {year} {2021})}\BibitemShut {NoStop}%
\bibitem [{\citenamefont {Rogers}\ and\ \citenamefont {Peiris}(2021)}]{rogers2021strong}%
  \BibitemOpen
  \bibfield  {author} {\bibinfo {author} {\bibfnamefont {K.~K.}\ \bibnamefont {Rogers}}\ and\ \bibinfo {author} {\bibfnamefont {H.~V.}\ \bibnamefont {Peiris}},\ }\href@noop {} {\bibfield  {journal} {\bibinfo  {journal} {Physical Review Letters}\ }\textbf {\bibinfo {volume} {126}},\ \bibinfo {pages} {071302} (\bibinfo {year} {2021})}\BibitemShut {NoStop}%
\bibitem [{\citenamefont {Dalal}\ and\ \citenamefont {Kravtsov}(2022)}]{dalal2022excluding}%
  \BibitemOpen
  \bibfield  {author} {\bibinfo {author} {\bibfnamefont {N.}~\bibnamefont {Dalal}}\ and\ \bibinfo {author} {\bibfnamefont {A.}~\bibnamefont {Kravtsov}},\ }\href@noop {} {\bibfield  {journal} {\bibinfo  {journal} {Physical Review D}\ }\textbf {\bibinfo {volume} {106}},\ \bibinfo {pages} {063517} (\bibinfo {year} {2022})}\BibitemShut {NoStop}%
\bibitem [{\citenamefont {Lian}\ \emph {et~al.}(2022)\citenamefont {Lian}, \citenamefont {Zasowski}, \citenamefont {Hasselquist}, \citenamefont {Holtzman}, \citenamefont {Boardman}, \citenamefont {Cunha}, \citenamefont {Fern{\'a}ndez-Trincado}, \citenamefont {Frinchaboy}, \citenamefont {Garcia-Hernandez}, \citenamefont {Nitschelm} \emph {et~al.}}]{lian2022quantifying}%
  \BibitemOpen
  \bibfield  {author} {\bibinfo {author} {\bibfnamefont {J.}~\bibnamefont {Lian}}, \bibinfo {author} {\bibfnamefont {G.}~\bibnamefont {Zasowski}}, \bibinfo {author} {\bibfnamefont {S.}~\bibnamefont {Hasselquist}}, \bibinfo {author} {\bibfnamefont {J.~A.}\ \bibnamefont {Holtzman}}, \bibinfo {author} {\bibfnamefont {N.}~\bibnamefont {Boardman}}, \bibinfo {author} {\bibfnamefont {K.}~\bibnamefont {Cunha}}, \bibinfo {author} {\bibfnamefont {J.~G.}\ \bibnamefont {Fern{\'a}ndez-Trincado}}, \bibinfo {author} {\bibfnamefont {P.~M.}\ \bibnamefont {Frinchaboy}}, \bibinfo {author} {\bibfnamefont {D.}~\bibnamefont {Garcia-Hernandez}}, \bibinfo {author} {\bibfnamefont {C.}~\bibnamefont {Nitschelm}},  \emph {et~al.},\ }\href@noop {} {\bibfield  {journal} {\bibinfo  {journal} {Monthly Notices of the Royal Astronomical Society}\ }\textbf {\bibinfo {volume} {511}},\ \bibinfo {pages} {5639} (\bibinfo {year} {2022})}\BibitemShut {NoStop}%
\bibitem [{\citenamefont {Magrini}\ \emph {et~al.}(2016)\citenamefont {Magrini}, \citenamefont {Coccato}, \citenamefont {Stanghellini}, \citenamefont {Casasola},\ and\ \citenamefont {Galli}}]{magrini2016metallicity}%
  \BibitemOpen
  \bibfield  {author} {\bibinfo {author} {\bibfnamefont {L.}~\bibnamefont {Magrini}}, \bibinfo {author} {\bibfnamefont {L.}~\bibnamefont {Coccato}}, \bibinfo {author} {\bibfnamefont {L.}~\bibnamefont {Stanghellini}}, \bibinfo {author} {\bibfnamefont {V.}~\bibnamefont {Casasola}}, \ and\ \bibinfo {author} {\bibfnamefont {D.}~\bibnamefont {Galli}},\ }\href@noop {} {\bibfield  {journal} {\bibinfo  {journal} {Astronomy \& Astrophysics}\ }\textbf {\bibinfo {volume} {588}},\ \bibinfo {pages} {A91} (\bibinfo {year} {2016})}\BibitemShut {NoStop}%
\end{thebibliography}%

\end{document}